\documentclass[conference,a4paper]{APSIPA2021}
\usepackage{multirow}
\usepackage{graphicx}
\usepackage{amsmath}
\usepackage[psamsfonts]{amssymb}
\usepackage{amsxtra}
\usepackage{cite}
\usepackage{threeparttable}
\usepackage{subfigure}
\usepackage{float}

\begin{document}
	
	\title{Source Tracing: Detecting Voice Spoofing}
	
	\author{%
		\authorblockN{%
			Tinglong Zhu\authorrefmark{1}, Xingming Wang\authorrefmark{1}\authorrefmark{2}, Xiaoyi Qin\authorrefmark{1}\authorrefmark{2}, Ming Li\authorrefmark{1}\authorrefmark{2}
		}
		\authorblockA{%
			\authorrefmark{1}
			Data Science Research Center, Duke Kunshan University, Jiangsu, China \\
		}
		\authorblockA{%
			\authorrefmark{2}
			School of Computer Science, Wuhan University, Wuhan, China\\
			E-mail: tinglong.zhu@alumni.duke.edu xingming.wang@dukekunshan.edu.cn\\xiaoyi.qin@dukekunshan.edu.cn ming.li369@duke.edu}
	}

	\maketitle
	\thispagestyle{empty}
	
	\begin{abstract}
		Recent anti-spoofing systems focus on spoofing detection, where the task is only to determine whether the test audio is fake. However, there are few studies putting attention to identifying the methods of generating fake speech. Common spoofing attack algorithms in the logical access (LA) scenario, such as voice conversion and speech synthesis, can be divided into several stages: input processing, conversion, waveform generation, etc. In this work, we propose a system for classifying different spoofing attributes, representing characteristics of different modules in the whole pipeline. Classifying attributes for the spoofing attack other than determining the whole spoofing pipeline can make the system more robust when encountering complex combinations of different modules at different stages. In addition, our system can also be used as an auxiliary system for anti-spoofing against unseen spoofing methods. The experiments are conducted on ASVspoof 2019 LA data set and the proposed method achieved a 20\% relative improvement against conventional binary spoof detection methods.
	\end{abstract}
	
	\section{Introduction}
	\begin{table*}[!htbp]
		\caption{Partial of Summary of LA spoofing systems\cite{wang2020asvspoof}. * indicates neural networks.}
		\center
		\label{tab:sp_comp}
		\begin{tabular}{|l|l|l|l|l|l|l|l|l|}
			\hline
			& Input          & Input processor & Duration   & Conversion       & Speaker representation & Waveform generator       &Usage 			\\ \hline
			A01 & Text           & NLP             & HMM        & AR RNN*          & VAE*                   & WaveNet*                 & Eval              \\
			A02 & Text           & NLP             & HMM        & AR RNN*          & VAE*                   & WORLD                    & Train            \\
			A03 & Text           & NLP             & FF*        & FF*              & One hot embed.         & WORLD                    & Train             \\
			A04 & Text           & NLP             & -          & CART             & -                      & Waveform concat.         & Train             \\
			A05 & Speech (human) & WORLD           & -          & VAE*             & One hot embed.         & WORLD                    & Eval              \\
			A06 & Speech (human) & LPCC/MFCC        & -          & GMM-UBM          & -                      & Spectral filtering + OLA & Train             \\ \hline
			A07 & Text           & NLP             & RNN*       & RNN*             & One hot embed.         & WORLD                    & Eval        \\ 
			A08 & Text           & NLP             & HMM        & AR RNN*          & One hot embed.         & Neural source-filter*    & Train             \\
			A09 & Text           & NLP             & RNN*       & RNN*             & One hot embed.         & Vocaine                  & Train             \\
			A10 & Text           & CNN+bi-RNN*     & Attention* & AR RNN+CNN*      & d-vector (RNN)*        & WaveRNN*                 & Train             \\
			A11 & Text           & CNN+bi-RNN*     & Attention* & AR RNN+CNN*      & d-vector (RNN)*        & Griffin-Lim              & Train             \\
			A12 & Text           & NLP             & RNN*       & RNN*             & One hot embed.         & WaveNet*                 & Train             \\
			A13 & Speech (TTS)   & WORLD           & DTW        & Moment matching* & -                      & Waveform filtering       & Train            \\
			A14 & Speech (TTS)   & ASR*            & -          & RNN*             & -                      & STRAIGHT                 & Train             \\
			A15 & Speech (TTS)   & ASR*            & -          & RNN*             & -                      & WaveNet*                 & Train             \\
			A16 & Text           & NLP             & -          & CART             & -                      & Waveform concat.         & Train             \\
			A17 & Speech (human) & WORLD           & -          & VAE*             & One hot embed.         & Waveform filtering       & Train             \\
			A18 & Speech (human) & MFCC/i-vector   & -          & Linear           & PLDA                   & MFCC vocoder             & Train             \\
			A19 & Speech (human) & LPCC/MFCC       & -          & GMM-UBM          & -                      & Spectral filtering+OLA   & Train             \\ \hline
			
		\end{tabular}
	\end{table*}
	
	The performance of both text-to-speech (TTS) and voice conversion (VC) systems has improved significantly in the past few years with the significant development of deep learning \cite{tan2021survey} and advanced training strategies\cite{yue2022exploring}. Many notable high-performance TTS and VC systems includes Tacotron systems\cite{wang2017tacotron,shen2018natural}, Fast Speech systems\cite{ren2019fastspeech,ren2020fastspeech}, VITS system\cite{kim2021conditional}, DelightfulTTS\cite{liu2022delightfultts}, etc. are proposed recently. This has exposed human users and ASV systems to increasingly serious potential attack threats and security concerns\cite{evans2013spoofing}. Therefore, building a trustworthy audio anti-spoofing system gradually attracts more and more attention.
	The most well-known fake audio detection challenge, the Automatic Speaker Verification and Spoofing Countermeasures (ASVspoof) challenge \cite{yamagishi2021asvspoof,nautsch2021asvspoof}, has been held since 2013 and focuses on building an audio spoofing countermeasure (CM) system for the ASV system. TTS/VS synthesized speech is often considered as the logical access (LA) attack. Generally, a CM system consists of a front-end feature extractor and a back-end classifier. The feature extractors typically extract handcrafted acoustic features based on the original waveform. Many acoustic features such as Constant Q Cepstral Coefficient (CQCC) \cite{todisco2016new}, Group Delay Gram (GD Gram) \cite{tom2018end}, Joint Gram \cite{cai2019dku} and Inverted Mel-Frequency Cepstral Coefficients (IMFCC) \cite{chakroborty2009improved} have been shown to be useful for audio anti-spoofing task. The back-end classifier usually identifies whether audio is a spoof or not based on the extracted features. More and more deep learning-based models and loss functions have been proposed to achieve better performance. In the latest ASVspoof2021 challenge, Tomilov et al. uses an LCNN-based \cite{wu2018light} architecture and achieved impressive performance \cite{tomilov2021stc}.
	Despite there are many works on the CM system architecture, research on the problem of fake audio algorithm attributes analysis is relatively limited. Zhao et al. \cite{zhao2022multi} uses the multi-task learning strategy to add the classification of known spoofing approaches to the existing CM framework and achieved noticeable performance improvement. However, this work only has an overall concept of detecting a set of systems and does not subdivide them according to different attributes at different levels. Therefore, this setup cannot handle unseen attacking scenarios. A similar solution was used by Borrelli et al. \cite{borrelli2021synthetic} to include a fake method detection module in a CM system using the multi-task learning approach. The unseen scenarios are considered as an open set classification problem. However, the approaches for generation are used for spoofing method classification.
	
	In the field of deep forgery image and video detection, the problem of traceability about forgery algorithms has also attracted great attention in recent years. Jain et al. \cite{jain2021improving} uses six categories of face forgery algorithm labels as training targets for a forgery recognition system. In the testing phase, the authors achieved better generalization performance than a simple binary classification system by fusing all forgery algorithm categories and treating the system as a binary spoofing detection system.
	
	TTS and VC systems can be divided into components such as speaker represent, waveform generator, and front-end models that convert text to a sequence of linguistic features \cite{wang2020asvspoof}. An arbitrary spoofing system can be constructed by combining different components, making the CM system for LA access more challenging
	
	In this work, we propose a framework for detecting spoofing attributes using multi-task learning to deal with the spoofing systems constructed by different combinations of TTS and VC modules. In other words, we want to use our framework to trace the attributes of an arbitrary speech synthesis or conversion system and determine what kind of algorithm is used during different stages. This could help to detecting unseen spoofing systems with one or more known attributes. For our work, we trace the following attributes of the complex LA spoofing systems:
	
	\begin{itemize}
		\item  Conversion
		
		\item  Speaker representation
		
		\item  Waveform generator
	\end{itemize}

	We re-partition the training set and evaluation set of the ASVspoof 2019 dataset\cite{wang2020asvspoof} of the LA task for out experiments. Under our multi-task training strategy, we achieve 88.4\% accuracy in identifying conversion algorithms, 51.5\% accuracy in detecting speaker representation modules (acoustic models, speaker encoder, etc.) and 77.5\% accuracy in identifying waveform generator by a single model. Moreover, our system can also be used as an auxiliary system for anti-spoofing detection to achieve better performance against unseen spoofing systems.
	
	\section{Proposed Method}
	\subsection{Related Works}
	The existing work for tracing the spoofing methods are mostly be used to improve the performance of the anti-spoof detection system rather than classifying the synthesis methods.  Li et al., \cite{li2019anti} identity the spoofing algorithm as an additional task under a multi-task training framework to improve the system performance on top of the anti-spoof countermeasure task. But the generating methods of test audio are seen in the training set. Borrelli et al., \cite{borrelli2021synthetic} conduct research on detecting and classifying spoofing methods. However, neither of them have studied  the attribute classification for spoofing attacks,  \cite{borrelli2021synthetic} only classifies different spoofing methods based on the waveform generator, while Li et al., \cite{li2019anti} detects the whole fake system without further subdivision. Adopting a set of attributes to describe the spoofing methods at different stages of the whole pipeline could enhance the robustness of the current spoofing method identification when detecting unseen spoofing methods. In order to improve the spoof detecting system's robustness towards those spoofing systems that are not directly included in the training set, but part of their modules are similar to the ones of other spoofing systems in the training set, we here propose a multi-task attribute classification training strategy. In this paper, we focus on classifying attributes of the spoofing systems. 
	
	%
	%
	%
	\subsection{Multi-Label Classification} 
	As shown in the training part of Fig. \ref{fig:com_mul_bin}(b)), according to the generating pipeline of spoofing speech, there are three attributes that are most important: Conversion, Speaker Representation, and Waveform Generator.
	\subsubsection{Conversion}
	Here the conversion denotes the feature transformation modules. The goal is to convert the input feature to match the target speaker's voice.
	
	\subsubsection{Speaker Representation}
	Attackers can take advantage of speaker representation to imitate a target speaker's voice. This may include speakers registered in the ASV security systems. Typically speaker representation is a high-dimensional vector that contains the speaker's embedding or index in the training data, timbre, etc. With speaker representation, attackers are able to generate speech according to  target speaker's characteristics using TTS or VC algorithms.
	
	\subsubsection{Waveform Generator}
	Waveform generator performs the conversion from acoustic features to the corresponding speech signals which are also called vocoders. The performance of waveform generator is highly correlated to the quality of the synthesized speech.
	
	\subsection{Spoofing Attribute Classification}
	We propose a training strategy that different back-end classifiers share the same front-end model.
	\begin{equation}
		\mathbf{e}_{i} = Z(X_{i})
	\end{equation}
	Where $\mathbf{e}_{i}\in \mathcal{\mathbf{R}}^d$, indicates the output vector extracted by front-end model $z(.)$ from $i_{th}$ audio. In this work, we define three spoofing attributes detection classifiers mentioned above: conversion, speaker representation, and waveform generator. The loss functions of classifiers are:
	
	\begin{gather}
		l_{conv} = L_{CE}(C_{conv}(e_{i}),y^{conv}_{i}) \\
		l_{spk} = L_{CE}(C_{spk}(e_{i}),y^{spk}_{i}) \\
		l_{wg} = L_{CE}(C_{wg}(e_{i}),y^{wg}_{i}) 
	\end{gather}
	
	Where  $conv$ denotes conversion, $wg$ denotes waveform generator and $spk$ denotes speaker representation and $C_{conv}$,$C_{spk}$ and $C_{wg} $ represent the corresponding attribute classifiers. And $y_{i}^{class}$ denotes the predicted label for each corresponding attribute. We apply the Cross-Entropy loss as our loss function. The final loss is formulated as a weighted summation and in which $\lambda_{i}$ is the weight value for different attributes:

	\begin{equation}
		l_{total} = \lambda_{1}l_{conv}+\lambda_{2}l_{spk}+\lambda_{3}l_{wg}
	\end{equation}

	\begin{figure*}[htbp]
		
		\centering
		\subfigure[]
		{
			\centering
			\includegraphics[scale=0.24]{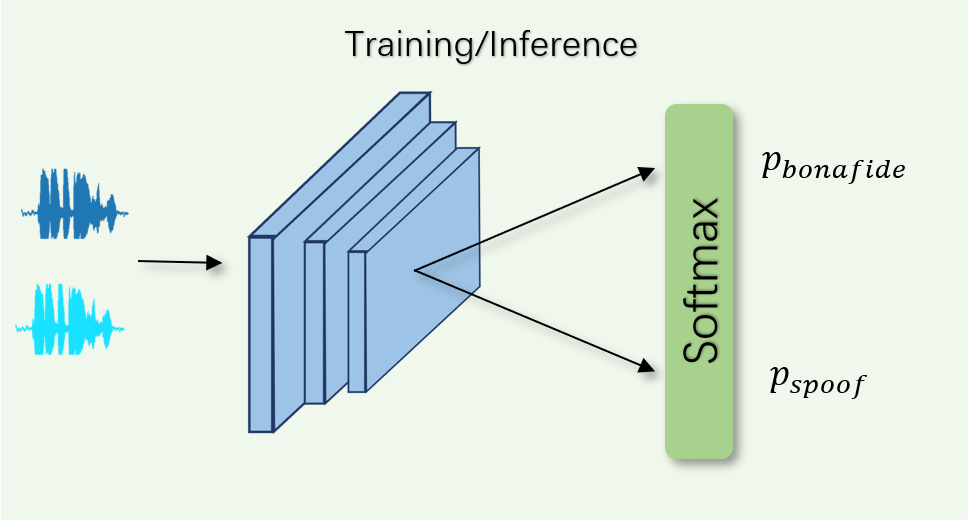}
		}
		\subfigure[]
		{
			\centering
			\includegraphics[scale=0.24]{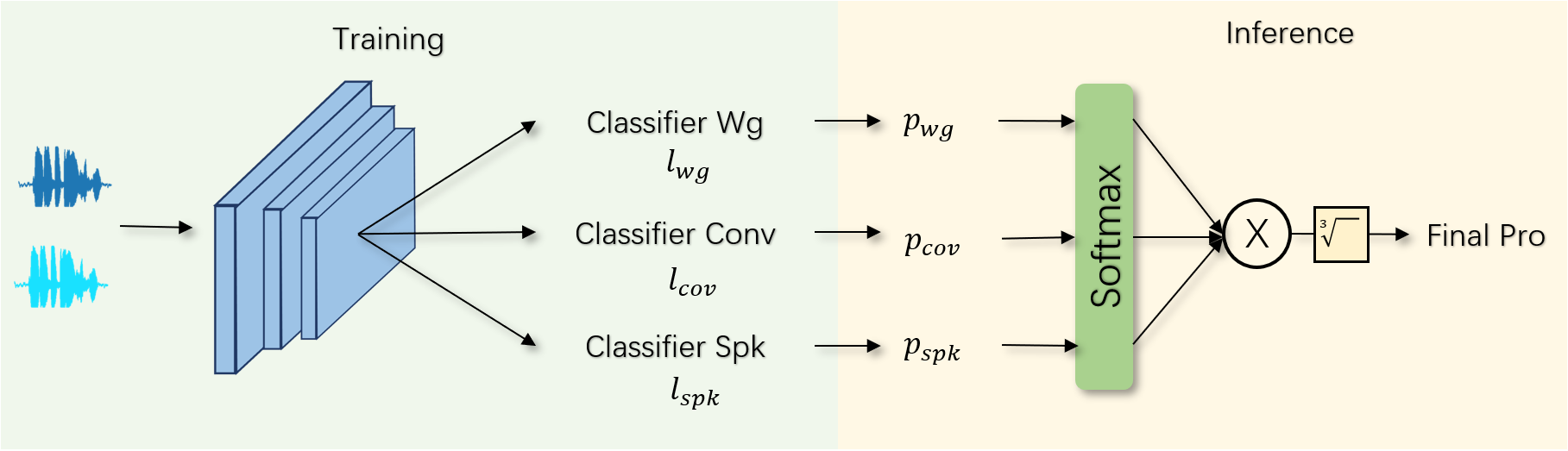}
		}
		\caption{Comparison between conventional spoof detection system and multi-task learning based spoof detection system }
		\label{fig:com_mul_bin}
	\end{figure*}

	\subsection{Anti-spoofing Countermeasure}
	\label{sec:imple}
	In addition to being able to trace the attribute of the spoofing system, the proposed method is also able to improve the performance of the anti-spoofing countermeasure system. Unlike most CM systems (Fig. \ref{fig:com_mul_bin}(a)), which only has one classifier for determining whether the input speech utterance is spoofed, our system has three classifiers for spoofing detection. This enables us to combine the bona fide probability from each classifier to get the final probability of whether the it is spoofed or not (Fig. \ref{fig:com_mul_bin}(b)). We adopt the cubic root of the product after multiplying three spoof probabilities from the classifiers as the final spoof score. 
	
	\begin{equation}
		s_{spoof} = \sqrt[3]{p_{spoof_{conv}} \times p_{spoof_{wg}}\times p_{spoof_{spk}}} 
	\end{equation}
	
	\section{Experiment Setup}
	\subsection{Dataset Division}
	\label{sec:data_usage}
	In our experiments, we use the ASVSpoof 2019 LA task's dataset\cite{wang2020asvspoof}, which has 121461 utterances in total. As shown in Table. \ref{tab:sp_comp}, the methods used in the original evaluation set and development set (A07-A19) are not fully covered in the original training set (A01-A06). Thus we reconstruct\footnote{https://github.com/kurisujhin/anti-spoof-source-tracing} the training set and evaluation sets. To make sure the attributes of all the methods in the evaluation are covered in the training set and there is no speaker overlap between these sets, we first divided all speakers in the LA dataset to two parts, one as training speaker set and the other for evaluation speaker set. For our evaluation set ,we choose the utterances from our evaluation set speakers with label bona fide, A01, A05, and A07 to form our evaluation set. And we select utterances with label A02-A04, A06, A08-A19 from our training set speakers and these speakers' bona fide utterances to form our training set. Note that the division will inevitably leave some utterances unused in the experiment. In this setup, our training set contains 67 speakers and 79620 utterances, while the evaluation set has 11 speakers and 5832 utterances. 
	\subsection{Label Assignment }

	As mentioned above, we proposed tracing the spooﬁng system's attributes. There are three attribute labels to train our classiﬁers, which means each utterance has three labels. Table.II presents the detail of the label assignment. To make the model more generalized, we divided the waveform generator into NN-based and non-NN-based methods. In addition, we combined all RNN-related methods in conversion attributes to one label. We keep the original labeling with \cite{wang2020asvspoof} for the speaker representation attribute.

	\begin{table}[!htbp]
		\small
		\centering
		\caption{Labeling for each attribute}
		\begin{tabular}{|c|c|}
			\hline
			Attribute   (Classifier)            & Methods                    \\ \hline
			\multirow{3}{*}{Waveform generator} & Nerual Network methods     \\ \cline{2-2} 
			& non Neural Network methods \\ \cline{2-2} 
			& bona fide                  \\ \hline
			\multirow{5}{*}{Speaker represent}  & VAE                        \\ \cline{2-2} 
			& One hot embed.              \\ \cline{2-2} 
			& d-vector (RNN)             \\ \cline{2-2} 
			& PLDA                       \\ \cline{2-2} 
			& bona fide                  \\ \hline
			\multirow{7}{*}{Conversion}         & RNN related methods        \\ \cline{2-2} 
			& FF                         \\ \cline{2-2} 
			& CART                       \\ \cline{2-2} 
			& VAE                        \\ \cline{2-2} 
			& GMM-UBM                    \\ \cline{2-2} 
			& Moment matching            \\ \cline{2-2} 
			& Linear                     \\ \cline{2-2} 
			& bona fide                  \\ \hline
		\end{tabular}
		\label{tab:lab}
	\end{table}

	\subsection{Models}
	We validate our strategies using two models, where one is based on ResNet34\cite{he2016deep}, and the other is based on the RawNet2\cite{tak2021end}. For the ResNet34 based model, we adopt the model structure of the ASV system proposed by Cai et al.\cite{cai2018exploring}. We apply log-FBank algorithm with 80-dimension Mels for feature extraction for the ResNet34 systems. For the RawNet based model, we employ the SincNet\cite{ravanelli2018speaker} based RawNet2 system\cite{tak2021end} in the experiment. We use the log-FBank features for the ResNet34 model, while for the RawNet2 model, we directly input the truncated/concatenated signals to the model. In a single experiment, all three classifiers receive the same inputs from the same front-end model.

	\section{Experiment Results}
	
	\subsection{Spoofing attribute classification}
	
	The performance overview of the ResNet34 and RawNet2 multi-task system are given in Table. \ref{tab:exp1}. We report each attribute's accuracy and clearly observe that the recognition accuracies of Conversion and Waveform generator attributes are over 80\%. But on the other hand, the accuracy of speaker representation is only about 50\%. This is due to the fact that the speaker representation is a latent vector built by the conversion model and not explicitly expressed on the signal.
	
	\begin{figure}[htbp]
		
		\centering
		\includegraphics[width=\linewidth]{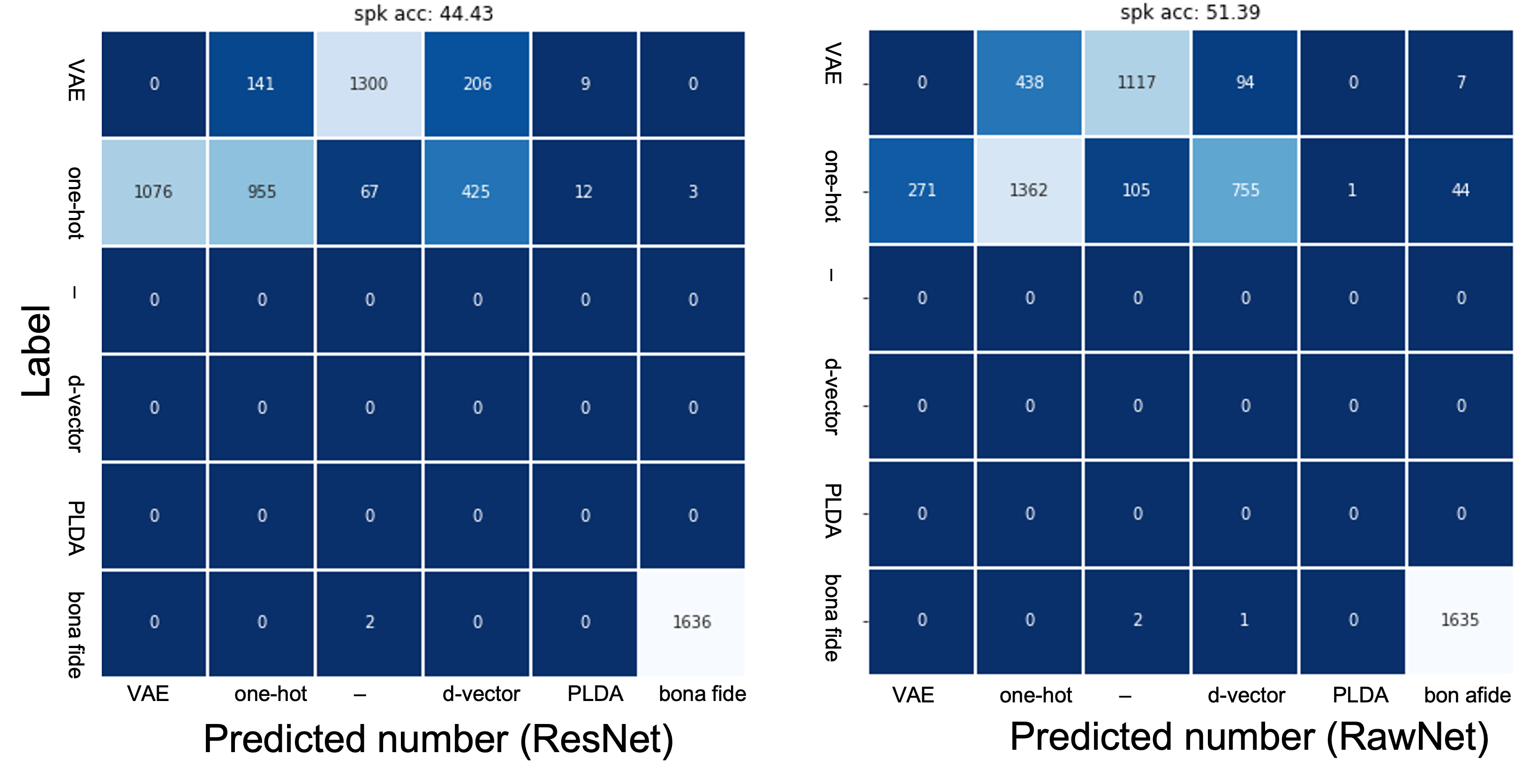}
		\caption{Predicted result of different model under the speaker represent attribute}
		\label{fig:spk_acc}
		
	\end{figure}
	Therefore, we further analyze and count the predicted results of speaker represent as shown in Fig. \ref{fig:spk_acc}. The results presents that although speaker represent attribute can not be correctly classified, the classifier still has a high accuracy in the discrimination of bonafide.

	\begin{table}[!htbp]
		\small
		\centering
		\caption{Multi-task Training on ASVSpoof 2019 LA dataset}
		\begin{tabular}{|l|c|c|c|}
			\hline
			Model    & Conv. Acc(\%) & Spk. Acc (\%) & Wg. Acc (\%) \\ \hline
			ResNet34 & 86.5               & 44.43                      & \bf{84.47}                       \\ \hline
			Rawnet 2 & \bf{88.41}         & \bf{51.46}                      & 77.54                       \\ \hline
		\end{tabular}
		\label{tab:exp1}
	\end{table}

	\subsection{Spoofing detection}
	
	As shown in Section. \ref{sec:imple}, our method could not only track the fake audio's attribute but also implement spoofing detection. Since the part of evaluation data is considered as training set, we recontribute the test set using new evaluation data mentioned in \ref{sec:data_usage}.  In this experiment, we adopt the conventional spoof detection system, which implement the binary classifier for spoofing detection to determine whether the test audio is spoofed, as baseline system, named Binary. The results presents in Table. \ref{tab:exp2}. Since training and evaluation set is small and the different training seed will slightly influence the results\cite{aasist}, all results reported in this paper are best result. 
	As presented in Table. \ref{tab:exp2}, our strategy can help improve the system performance by at least 20$\%$. ResNet-based model even achieves about 80\% relative improvement than baseline system. Since the original binary classification task (bona fide and spoof classification) is split to three attribute classifiers, the model space for the spoof detector has been increased, which may be a possible explanation for the performance improvement. 
	
	\begin{table}[!htbp]
		\small
		\centering
		\caption{Performance comparison between conventional spoof detection system and multi-task learning based spoof detection system}
		\begin{tabular}{|c|c|c|}
			\hline
			Model    & Method & EER [\%] \\ \hline
			ResNet34 & Multi-task (proposed) & \bf{0.012} \\ \hline
			ResNet34 & Binary & 0.066   \\ \hline
			RawNet2  & Multi-task (proposed) & \bf{0.187}   \\ \hline
			RawNet2  & Binary& 0.241   \\ \hline
		\end{tabular}
		\label{tab:exp2}
	\end{table}
	
	\section{Conclusions}
	In this paper, we show that our multi-task training strategy can help the model achieve acceptable performance in attribute source tracing for logical access spoofed utterances. The experiment results shows that our strategy can not only help source tracing but also help to improve the spoof detecting systems by combining the bona fide probabilities from three attribute classifiers. In addition to whether the utterance is spoofed or not, our training strategy can help to extract extra information like how the spoof system is designed, what algorithm the spoof systems used to generate a waveform, etc., which can help to improve the robustness of the spoof detection system towards those spoofing systems that are not directly included in the training set, but part of their modules are similar to the ones of other spoofing systems in the training set. Moreover, we present a new strategy to improve the performance of the spoof detecting system in addition to improving the front-end model, feature extraction, etc. For the future exploration of the source tracing, more spoofed data generated by different combinations of spoofing algorithms are needed.
	
	\section*{Acknowledgment}
	
	This research is funded in part by the National Natural Science Foundation of China (62171207), Science and Technology Program of Guangzhou City (202007030011) . Many thanks for the computational resource provided by the Advanced Computing East China Sub-Center.
	
	\bibliographystyle{IEEEtran}
	
	\bibliography{mybib}
\end{document}